\documentclass[12pt, a4paper]{article}
\usepackage{theorem}
\usepackage{makeidx}
\usepackage{amsmath}
\usepackage{amssymb}
\usepackage{epsfig}
\usepackage[latin1]{inputenc}
\usepackage{graphics} 
\begin{document}
\noindent
\renewcommand{\thefootnote}{\fnsymbol{footnote}}
\thispagestyle{empty}
\begin{center} {\huge{Socioeconomic Interaction and Swings in Business Confidence Indicators{\footnote{We would like 
to thank the Ifo-Institute, in particular Ms. S. Stallhofer, for providing us with data from the Ifo-Business-Climate survey. 
M.H. gratefully acknowledges financial support from DFG grant TR 120/12-1}}}}
\vspace{0,6cm}

\setcounter{footnote}{6}                                      

{{Martin Hohnisch, Sabine Pittnauer, Sorin Solomon and Dietrich Stauffer}}{\footnote{Address: Hohnisch: Research Group 
Hildenbrand, Department of Economics, University of Bonn,
Lenn\'estr. 37, D-53113 Bonn, Germany, and Department of Mathematics, University of Bielefeld, D-33501 Bielefeld, Germany 
(e-mail: Martin.Hohnisch@wiwi.uni-bonn.de);
Pittnauer: Research Group Hildenbrand and Institute of Public Economics, Department of Economics, University of Bonn,
Lenn\'estr. 37, D-53113 Bonn, Germany (e-mail: Sabine.Pittnauer@wiwi.uni-bonn.de);
Solomon: Racah Institute of Physics, The Hebrew University of Jerusalem, Givat Ram, Jerusalem 91904, Israel (e-mail: sorin@vms.huji.ac.il); 
Stauffer: Institute of Theoretical Physics, University of Cologne, Z\"ulpicher Str. 77, D-50923 K\"oln, Euroland (e-mail:
stauffer@thp.uni-koeln.de)}}

\end{center}
\vspace{0.0cm}
{\abstract{We propose a stochastic model of  interactive formation of individual expectations regarding the business 
climate in an industry. Our model is motivated by a business climate survey conducted since 1960 in Germany by
 the Ifo-institute (www.ifo.de). 
In accordance with the data structure of this survey, in our model there is associated 
to each economic agent (business manager) a random variable with a three-element state space representing 
her possible types of expectations. The evolution of individual expectations in a finite population is then modeled
 as a spatio-temporal stochastic process with local interaction between agents. 
An appropriate structure of the interaction between agents in our setting turns out
to be provided by a Festinger-function (in physics called energy function or Hamiltonian) of the Blume-Capel type.
Time series of the fractions of agents holding each type of expectations are
obtained for the model by Monte Carlo simulations.
We find that our model reproduces some generic features of the
empirical time series obtained from the German business-climate data, in particular the occurrence of abrupt large but rare swings. 
In our model, such swings occur as spontaneous phase changes between macroscopic states. }}

\newpage
\section{Introduction}
\renewcommand{\thefootnote}{\arabic{footnote}}
\setcounter{footnote}{0}                                      

The interactions-based approach to modeling socioeconomic behavior (for an overview, see \cite{DB}) is becoming increasingly popular among economists to explain collective phenomena such as herding behavior in financial markets. 
Blume and Durlauf define the interactions-based approach as ``focusing on direct interdependences between economic actors rather than those indirect interdependences which arise through the joint participation of economic actors in a set of markets.'' 

Our present paper focuses on a particular instance of socioeconomic interaction, whereby an economic agent tends
 to align herself with the opinions hold by others. Such group pressure effects on the opinion of individual group members have been
experimentally found and analysed in social
psychology \cite{A}\cite{F}. They have been explained within Leon Festinger's Social Comparison Theory as resulting
from a deep seeded need of a typical individual to have his/her opinions, thoughts, actions and attitudes compared with
others and to adapt oneself to the reference group if a substantial deviation from the group status is recognized.
In our particular context, we assume that a business manager is influenced in her expectations about the future business prospects in the industry to which her firm belongs by the expectation prevalent in her professional peer group. 

Socioeconomic interactions-based phenomena can be effectively modeled by the mathematical concept of a random
field{\footnote{The meaning 
attached to the term {\it{random field}} is different 
in mathematical physics and applied statistical physics. We follow in this paper the mathematical tradition to 
use the term random field for a family of random variables
indexed by a not linearly ordered set, such as the integer lattice $\mathbb{Z}^d$.}} (see \cite{Fo} for a pioneering contribution). In that stochastic approach in economic modeling,
to each agent is associated a random variable the state space of which is the set of possible individual
characteristics or decisions. The interaction between agents is then characterized by a family of conditional probabilities for an agent to assume some individual characteristic or decision given a fixed configuration of characteristics/decisions in the peer group of the agent.
The peer group structure is modeled by a certain graph structure on the set of agents. 

The particular structure of our model is motivated by a business climate survey performed monthly by the German Ifo institute.
 In that survey, business managers are asked to 
characterize their expectations about business prospects in their particular industry using the three categories,
 negative, neutral or positive. The time series obtained  from the survey data,{\footnote{The same
survey has been used in the  study by B. Flieth and J. Foster \cite{FF}, and, in fact, it was their study which attracted
our attention to the particular time series shown in Figure 1.}} depicted in Figure 1,
are used as an empirical benchmark  for our model.

Among the salient features in these
time series are sharp swings in the fractions of agents holding particular expectation types occurring at least four times between 1970 and 2000 and the tendency of the fractions to 
settle at a particular level in between the swings. 
We follow Flieth and Foster \cite{FF} in attributing such phenomena to herd behavior resulting
from the tendency of economic agents to 
align themselves with the prevalent expectation of peers.
However, in the present paper we use a micro-model with local instead of 
global interaction, the latter being used in \cite{FF}, to reproduce swings similar to those 
present in the empirical data.

In our model, large but rare swings in the time series of the fractions of agents holding particular expectation types occur spontaneously
possibly explaining similar swings in empirical time series for which no apparent explanation based on
 economically relevant events in the particular industry exists \cite{FF}. 
Such effects arise within our modeling framework as random collective coordination phenomena.

The structure of our paper is as follows. In Section 2 we present the basic specification of the spatio-temporal
stochastic process underlying our model. In Section 3 we present Monte Carlo generated time series of the fractions of 
agents holding  particular
expectation types. The paper concludes with a discussion of some aspects of our approach in Section 4.

\section{A model of expectation formation with local interaction between agents}

Let $A$ denote a finite set of economic agents. Given the specific context of our paper, we interpret $A$ as a population of business managers within a particular industry. To each agent $a\in A$ there is associated a variable $X_a$ with values from the set $S=\{-1, 0, 1\}$. 
The specification of a
three-element state space is motivated by the survey described in the introduction.
The realizations of $X_a$ are interpreted as expectation types hold by the agents regarding the business prospects
they expect to prevail in their industry over a certain time horizon. Those expectations are labeled
negative, neutral and positive and each is associated with an element from $S$. 

We assume socioeconomic interaction between agents in accordance with Festinger's theory of social comparison \cite{F}
resulting in agents tending to align themselves with the prevalent expectation type in their peer group.
The peer group of an agent $a\in A$ is identified with the set of next neighbors $\mathcal{N}(a)$ of agent $a$ with
respect to a graph structure imposed on $A$. For simplicity we assume the graph structure to be a finite square
sublattice $\Lambda$ of $\mathbb{Z}^2$ with periodic boundary conditions.

We specify an interaction pattern which seems to be implied by the phenomenological characteristics of the
empirical time series presented in Figure 1.
There are two types of expectations which seem to be prone to group pressure as the sharp swings in the
fraction of business managers holding one of these types appear to be a collective phenomenon.
In addition, there is a third type of expectation, which is more autonomous in the sense
that it appears to be less influenced by group pressure as no sharp swings occur.
 There are two macroscopic states
each one with characteristic population fractions of the non-autonomous expectation types. 
The autonomous expectation type coexists 
in both of the macroscopic states and is slightly correlated with the fraction of the non-autonomous types.

We obtain such an interaction pattern by the following specification of the local characteristics of the random field{\footnote{
See footnote 1}}.
First, the probability that an agent holding a non-autonomous expectation type changes to the
other non-autonomous type increases with the fraction of agents
in the peer group holding the other type and is almost one if all neighbors are of the other type. Second,
the probability that an agent holding  the autonomous expectation type changes to one of the
non-autonomous types is less dependent than in the first case on the fraction of agents in the peer group holding other types of expectation. 

For a quantitative formulation of the above specified interaction pattern, we define a mathematical object, 
which we call the {\it{Festinger-function}}{\footnote{The Festinger-function corresponds to the energy
function or Hamiltonian of a physical system.}}, $F: S^\Lambda\rightarrow \mathbb{R}$
 such that the probability of each configuration $x\in S^\Lambda$ can be written as
\begin{equation}
P(x) = \frac{1}{Z}e^{-F(x)}
\end{equation}
with the normalization constant \begin{equation} Z=\sum_{x \in S^\Lambda}e^{-F(x)}. \end{equation}
In this paper we confine ourselves to a generic  specification of the Festinger-function leaving a
 quantitative experimental investigation of such group pressure effects for future research (see Section 4). 
A concrete specification of the Festinger-function fulfilling the above requirements on the interaction structure is the following
\begin{equation}
F(x)= J \sum_{<i,j>}(x_i -x_j)^2
\end{equation}
with the sum over all pairs $i<j$ of next neighbors.
By that specification, the more neighboring agents have both assumed the same expectation type the higher is the probability of a 
given configuration because each such pair of neighboring agents contributes $0$ to the sum, 
while each pair of neighboring agents holding different expectation types
contributes either $J$ or $4J$ to the sum.
In effect, this generic specification captures the basic features of group pressure as specified above.




Next, we formulate a discrete-time stochastic dynamics for the atemporal model described in the previous paragraphs.
The type of stochastic dynamics we use is called Glauber dynamics \cite{LB}.  
Let $(X^{\tau})_{\tau\in\mathbb{N}}=(X^0, X^1, \dots)$ denote a family of random variables with state space $S^{\Lambda}$. 
The realization of the process at  time{\footnote{The time index  $\tau$ does not correspond to the time scale of the system. See remarks below in the main text.}} $\tau$ characterizing the configuration of all agents' expectations at $\tau$,
will be denoted by $x^{\tau}$.

The specification of the transition probabilities of the Glauber dynamics is as follows. 
At each point of  (computational) time $\tau \in \mathbb{N}$ only a single agent, denoted by $\hat{a}$, is allowed to 
change her expectation (To avoid notational clutter, we write $\hat{a}$ instead of $\hat{a}(\tau)$){\footnote{There
are various ways of selecting the agent $\hat{a}(\tau)$ at each computational step $\tau$. In our specification of the dynamics
we have chosen the ``type-writer'' algorithm, i.e. the agents are updated sequentially along the rows of the lattice.}}.
In a first step, one of the
two  expectation types other than that which the agent $\hat{a}$ currently holds is randomly independently selected
 with equal probability  assigned
to each of them. We denote that alternative expectation type for agent $\hat{a}$ as $\xi_{\tau+1}$. 
In a second step, the agent $\hat{a}$  might change to the expectation type $\xi_{\tau+1}$.
The probability of that event is denoted by $p_{\hat{a}, \tau}$ while
$1-p_{\hat{a}, \tau}$ denotes the probability of the complementary event that the agent $\hat{a}$ retains her expectation type, 
i.e. $x_{\hat{a}}^{\tau +1}=x_{\hat{a}}^{\tau}$.

The probability for agent $\hat{a}$ to assume at $\tau +1$ the expectation type $\xi_{\tau +1}$ is defined in the
 Glauber process as 
the following conditional probability 
\begin{equation}
p_{\hat{a}, \tau}=P(X_{\hat{a}}=\xi_{\tau +1}| \ (X_{\hat{a}}=\xi_{\tau +1}\vee X_{\hat{a}}=x_{\hat{a}}^{\tau}) \wedge
X_{\Lambda \setminus \hat{a}}=x_{\Lambda \setminus \hat{a}}^{\tau})
\end{equation}
with $P(\cdot)$ denoting the measure on the finite space of configurations $S^\Lambda$ associated with the
 atemporal model and derived 
from the Festinger-function according to  Equation 3.

Thus the transition probability $p_{a, \tau}$ obtains after some algebra to
\begin{equation}
p_{\hat{a}, \tau}=\frac{e^{-(F(x^{\tau +1})-F(x^{\tau}))}}{1+e^{-(F(x^{\tau +1})-F(x^{\tau}))}}
\end{equation}
In the above equation $x^{\tau +1}$ denotes the juxtaposition $(\xi_{\tau+1} \ x_{\Lambda\setminus \hat{a}})\in S^{\Lambda}$, i.e. 
the configuration in which $\xi_{\tau +1}$ replaces $x_{\hat{a}}^{\tau}$ in the configuration $x^{\tau}$.

Finally, we follow the usual concept to define a time step of the modelled system as the number of time steps in the Glauber process 
such that every agent is updated once.

The Glauber dynamics of a finite random field has the property that the distribution of the process at time $\tau$
 converges to the unique distribution $P$ of the atemporal model as $\tau \rightarrow \infty$ (see \cite{LB}).


\section{Results}
This section presents time series of the Glauber dynamics of our model described in the
previous section (along with a modified version of it) obtained by Monte Carlo (MC) 
simulations. The aim of our model is to reproduce some generic features of the 
empirical time series data of the fractions of agents holding particular expectation types depicted in Figure 1
(see also Figure 1 in Flieth and Foster \cite{FF}). In summary, these
features are, first, the occurrence of large but rare swings in the fractions of two
types of expectations which are hardly explainable by the factual situation of the
particular industry, second, the settling of the fractions on particular 
levels in between the transitions and, third, a relatively low fraction of the third type
of expectations slightly correlated with one of the other types of expectation.

We have used in the simulations a more general version of the Festinger-function than that specified in Equation 3. The 
Festinger-function can be written as 
\begin{equation}
F(x)= -2J \sum_{<i,j>}x_i x_j + \delta \sum_i x_i^2
\end{equation}
with $\delta=4J$. The parameter $\delta$, now taken as independent,
 allows us to control the fraction of the autonomous expectation type.

Figure 2 depicts a representative interval of MC-generated time series of the fractions
of expectation types associated with the state space elements $-1$ (top), $0$ (middle) and $1$ (bottom)
obtained from the specification in the Festinger-function with $\delta=0$ and $J=0.2976$.
 There are several spontaneous phase changes in which the prevalent fraction of expectations
 changes from $1$ to $-1$ or vice versa. 
It is known from previous work on the Blume-Capel model that with our choice of the Festinger-function
there are only two stable phases for 
$J>0$ and $\delta=0$, namely those in which either the element $-1$ or the element $1$ prevails. 

Moreover, it is known from work on similar models that the average length of the time 
period between two consecutive transitions from one macroscopic state to the other increases exponentially
with lattice size \cite{MOT}. The system stays on average
the same time in either of the two phases associated with a high fraction of $1$ or $-1$.

The fraction of the element $0$, representing the expectation type which is less influenced by 
other agents' expectations, does not correlate with the fractions of the other expectation types,
i.e. it is independent of the prevailing macroscopic state.
To account for the feature visible in Figure 1, namely a correlation of the 
fraction of the weakly interacting
expectation ($0$) to one of the other two, we have modified the basic model in that we have replaced the state space
$S=\{-1, 0, 1\}$ by $S'=\{-1.014, 0.05, 1\}$. The replacement of the element $0$ by the element $0.05$ makes
for an asymmetry in the model which results in reproducing the empirically observed correlations described above. In addition, the
element $-1$ is replaced by $-1.014$ to approximately equalize the minima of the Festinger-function to secure that the system
stays on average equally long in both of its macroscopic states. The resulting time series displaying the feature of 
a correlation between the autonomous expectation type with one of the non-autonomous expectation types is
presented in Figure 3. 

Finally, we remark that finite-size effects, which in fact underlie the results presented in this section, can 
be reasonably expected in some economic settings, in which the number of involved participants is by several magnitudes smaller than in 
similarly structured physical systems.

\section{Discussion}

In the presented simulation results, the constant characterizing the strength of interaction is set
 $\frac{1}{2J} = 1.68$, that is, near the value of critical temperature of the Blume-Capel model. Should we assume that business managers all know the Blume-Capel critical temperature? Of course, not. 
Instead, a simple heuristic argument can justify why the system is in the range of that value.
For $J$ much smaller than this critical value the variables associated with the various expectation types flip 
very fast; for $J$ much larger than this critical value nearly all such variables have the same value. 
Business, in contrast to a basket ball game, cannot survive if the manager changes opinion every second. Thus management moves 
towards slow changes in opinion, corresponding to $J$ near the critical point where relaxation times in infinitely large systems become infinite. 
On the other hand, managers are supposed to be reasonable such that positive and 
negative expectations are possible. If everybody would always follow the majority of neighbors (as in our model for large $J$), most companies would 
produce the same goods, offer the same services, and soon go bankrupt. 
Thus, $J$ moves automatically from too high values towards the range of the critical point where
all three types of expectations start to coexist. 

In our specification of the interaction structure by a Festinger-function of the Blume-Capel type, one
additional issue deserves further discussion. Does a frequent observation of an opinion which is, in a heuristic sense, 
``far away'' from one's own, exert more or less pressure to change one's opinion than of an opinion which
is ``closer'' to one's own?
The answer to this question seems to depend on the specific context. For example, individuals having strongly different
political opinions often refuse to talk to each other so that any influence is inhibited resulting in less pressure \cite{D}\cite{H}. On the other hand, if individuals are flexible enough to learn and respect the knowledge of peers, then presumably more pressure to change is exerted by opinions which are ``far away''.
We believe that in the context of our paper the second scenario is more appropriate and thus the chosen Festinger-function justified. 

It should be emphasized that a pattern of large but rare swings, characteristic for two of the time series
in the empirical data, can be obtained in the Glauber dynamics of a finite lattice model for many other
specifications of the Festinger-function/energy function. 
For example, we have obtained results very similar to those in Figure 2 (top) and (bottom) from 
the so-called three-state Potts model. 
However, the Potts model has two properties which make it in our modeling context less appropriate than the 
Blume-Capel model. First, the energy function in the Potts model is symmetric with respect to the elements
of the state space. There exist three symmetrical macroscopic states, in contrast to the empirical data in which
there is a strong asymmetry of the time series associated with the three types of expectations. Second, in the Glauber 
dynamics of the Potts model, transition probabilities violate a natural assumption related with the intuitive order on the set of expectation types.
For example, consider an agent holding a negative expectation in an environment with all agents holding positive
expectations. According to the Glauber dynamics, the agent will then switch to the positive expectation with high 
probability, will stick to the negative expectation with low probability and will switch to the neutral expectation
with an even lower probability. Clearly, the latter aspect of such a specification of the transition probabilities is counterintuitive.

A final remark addresses a methodological issue from the perspective of economics.
In our approach we do not pose the question why
individuals might behave in a conformistic manner. Instead a descriptive (statistical) approach is used in which the behavior
is described by conditional probabilities of an agent holding some expectation given a fixed configuration of expectations
of the other agents constituting her peer group. That description should be based on an appropriately designed
experiment, for instance,  in which individuals are asked to form an expectation faced with
a varying proportion of others in their peer group who allegedly hold some expectation. The proportion of agents holding 
some expectation conditional on the configuration of expectations in the peer group then corresponds to 
the local characteristics of the atemporal random field model.

\newpage
\begin{figure}[htp]
\begin{center}
\includegraphics[scale=0.8]{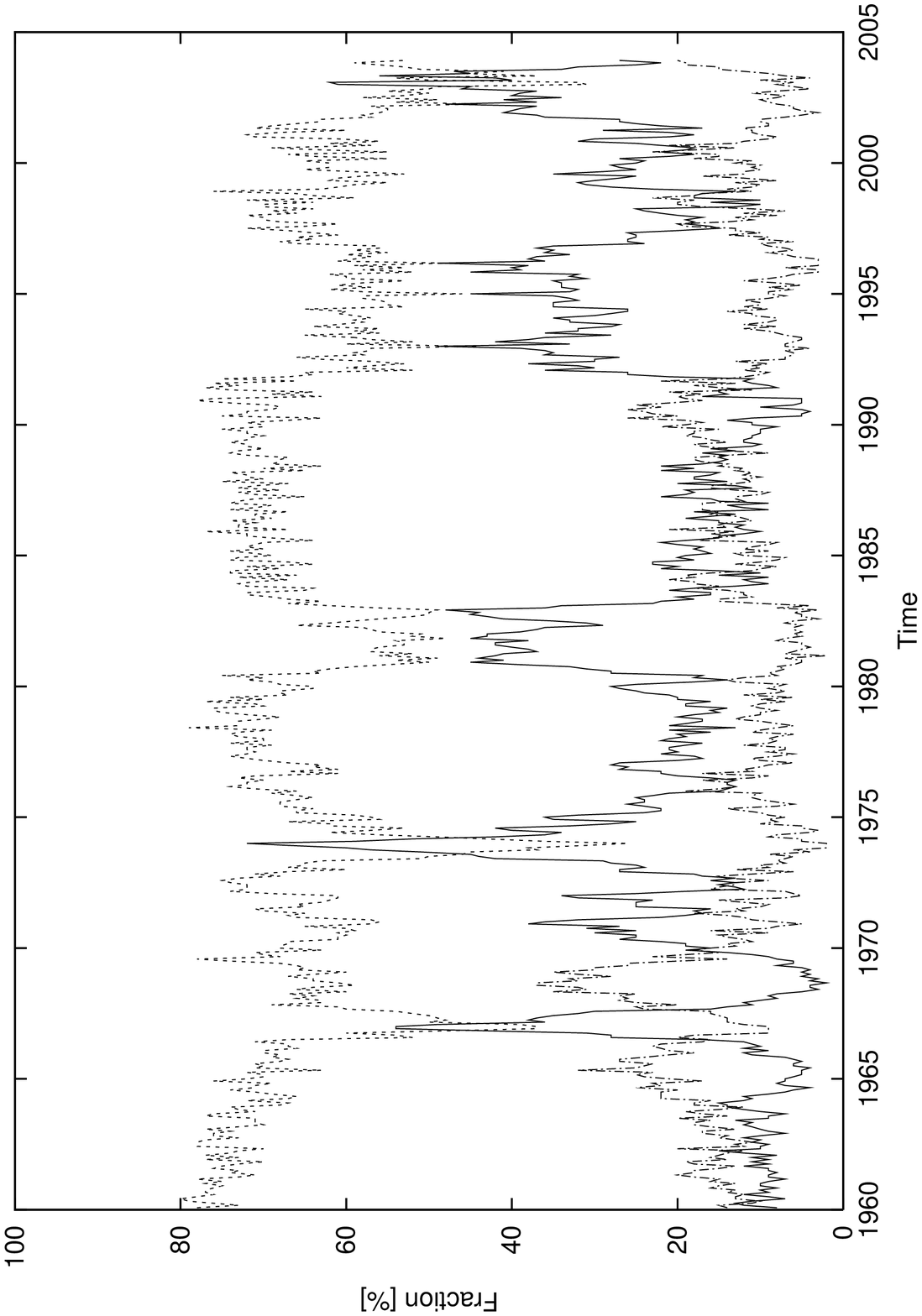}\\
\caption{\label{frequency1} Empirical time series of the fractions of particular expectation types; dashed line: neutral,
solid line: negative, dashed-dotted line: positive. Time series based on a slightly smaller data set 
have been previously published in \cite{FF}.}
\end{center}
\end{figure}

\begin{figure}[htp]
\begin{center}
\includegraphics[angle=-90,scale=0.38]{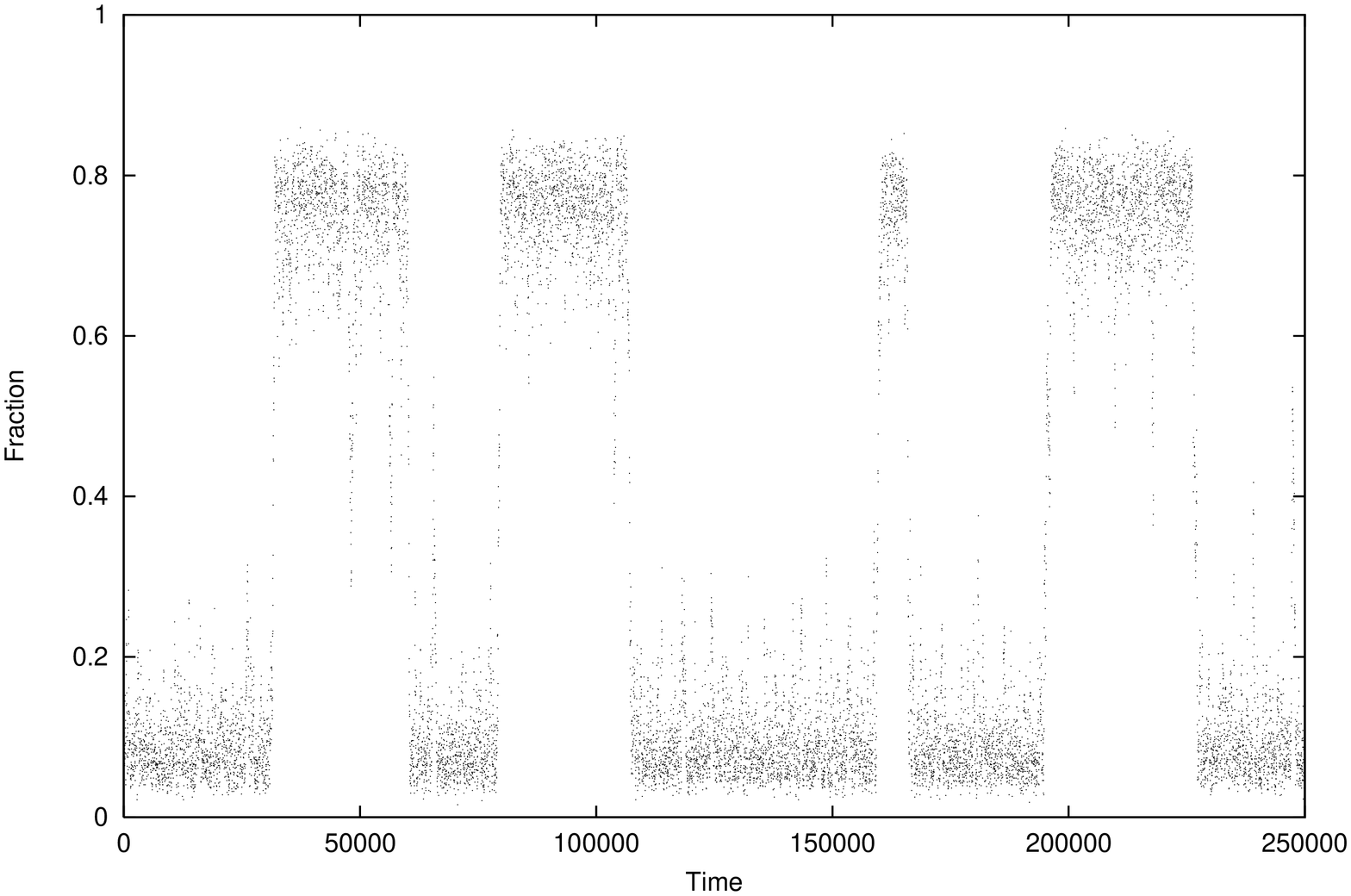}\\
\includegraphics [angle=-90,scale=0.38]{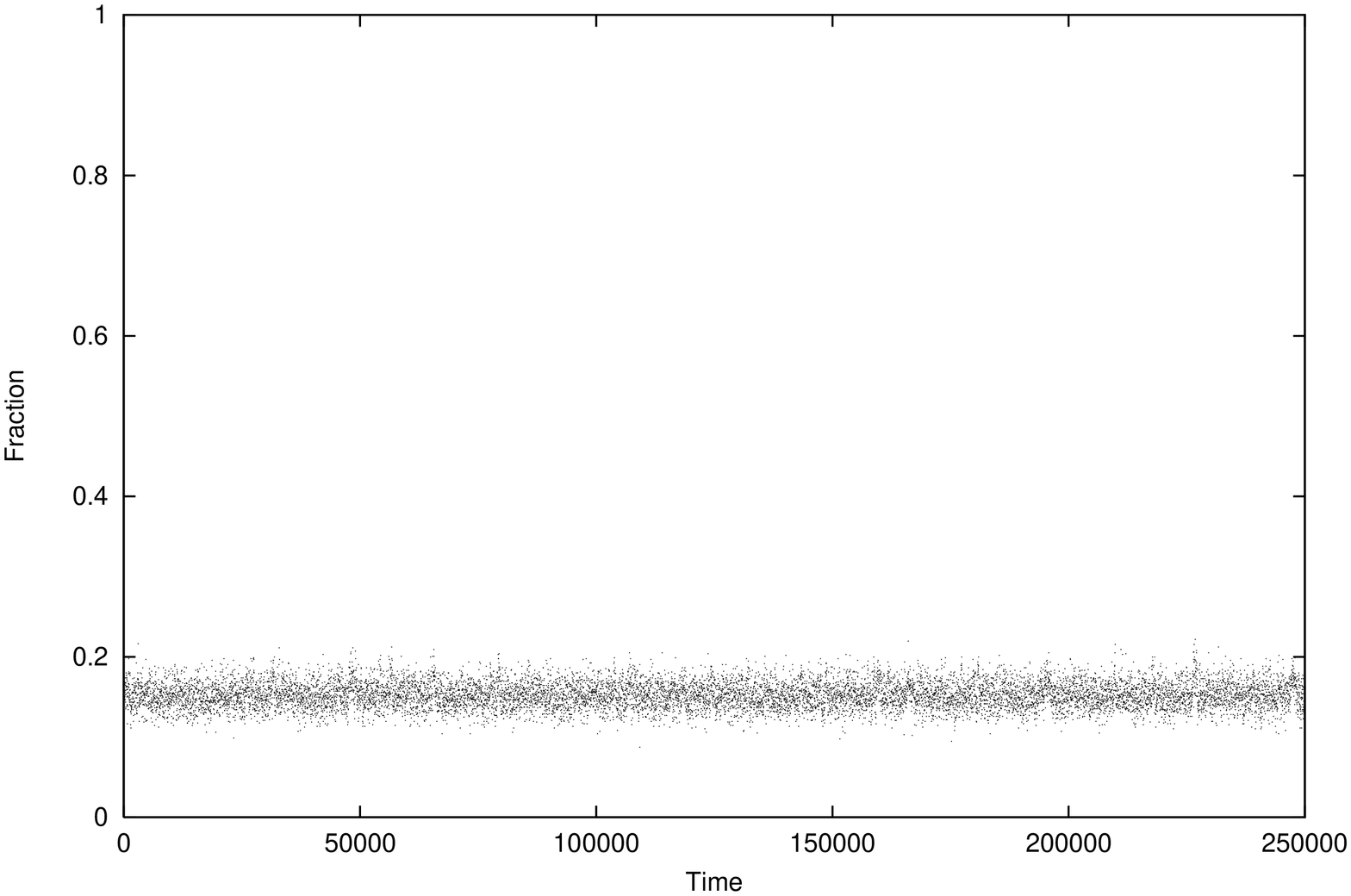}\\
\includegraphics[angle=-90,scale=0.38]{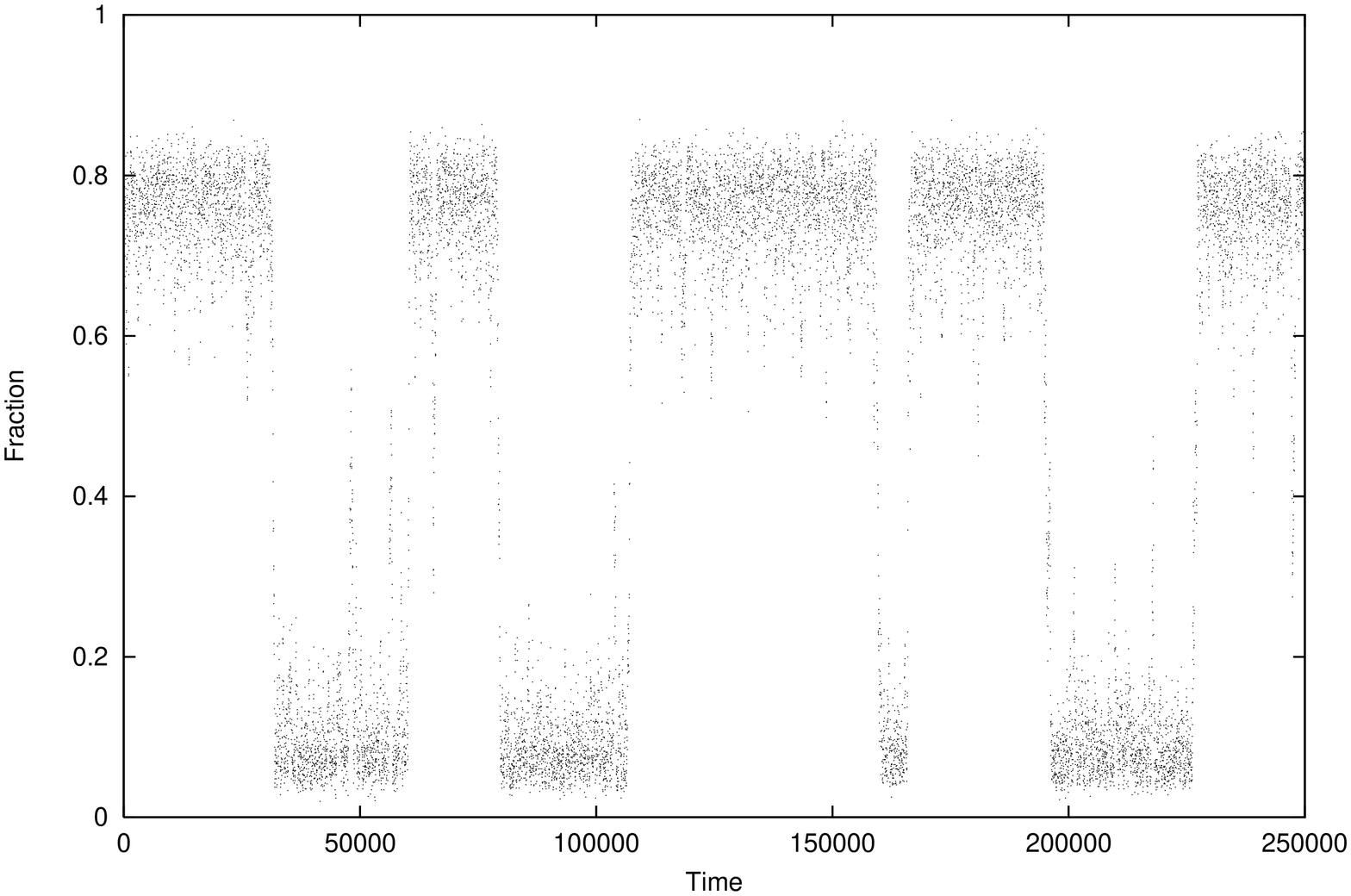}\\
\caption{\label{frequency2} Representative intervals of MC-generated time series of 
the fractions of expectation types associated with the state space elements $-1$ (top), $0$ (middle) and $1$ (bottom); the 
size of the population is 961 (31 $\times$ 31 square lattice), $\frac{1}{2J}=1.68$, $\delta=0$ }
\end{center}
\end{figure}

\begin{figure}[htp]
\begin{center}
\includegraphics[angle=-90,scale=0.38]{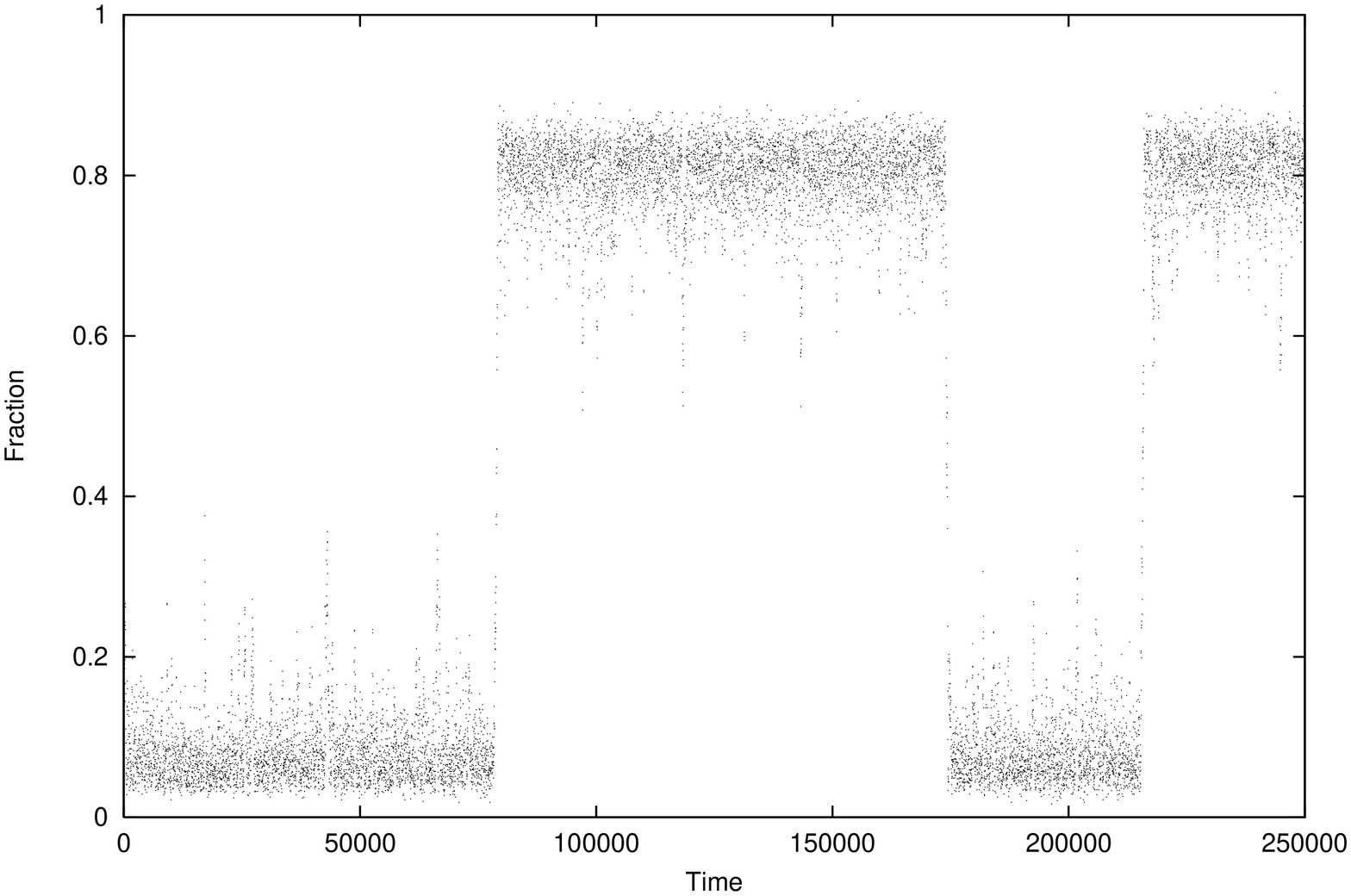}\\
\includegraphics[angle=-90,scale=0.38]{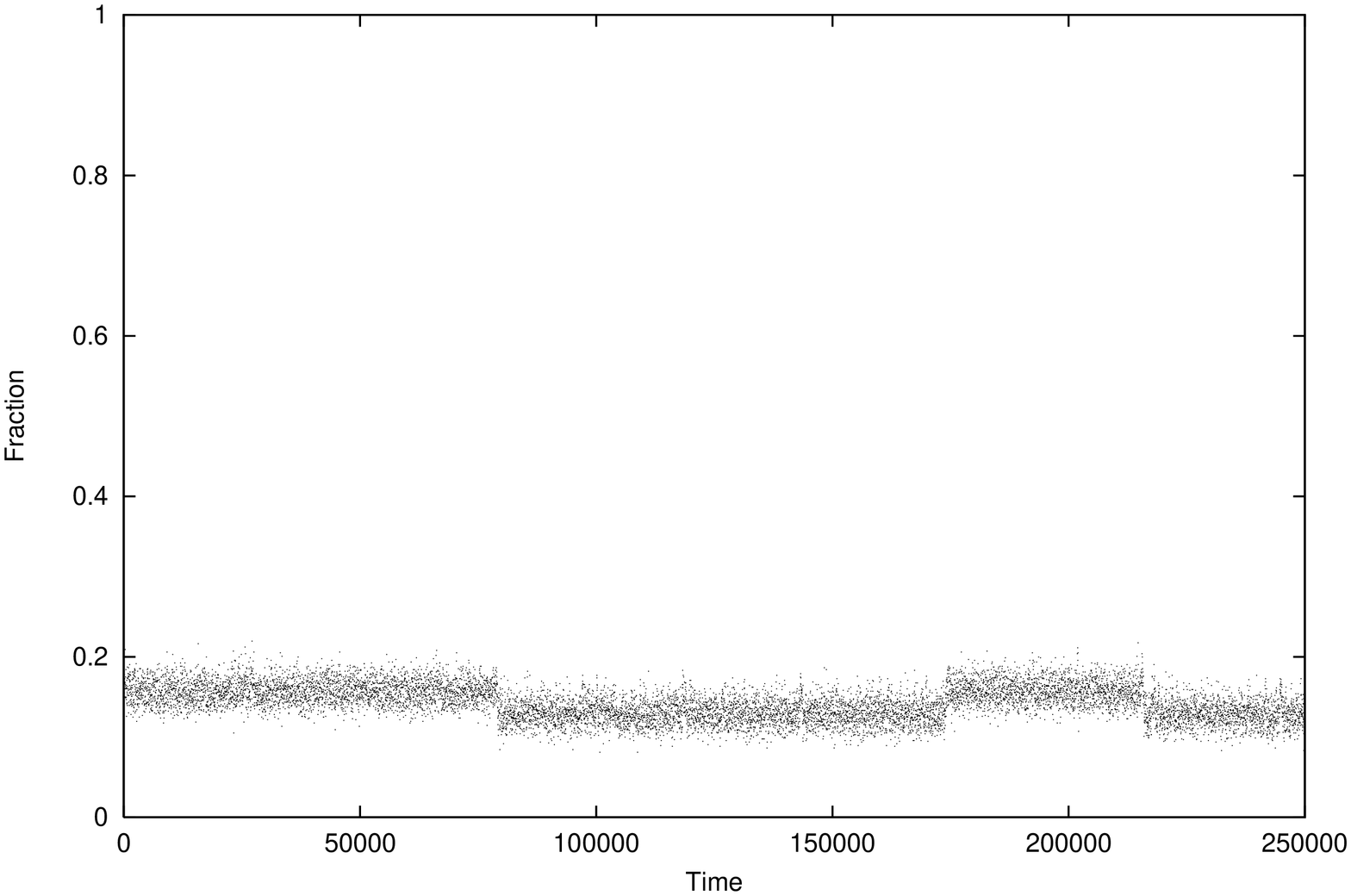}\\
\includegraphics[angle=-90,scale=0.38]{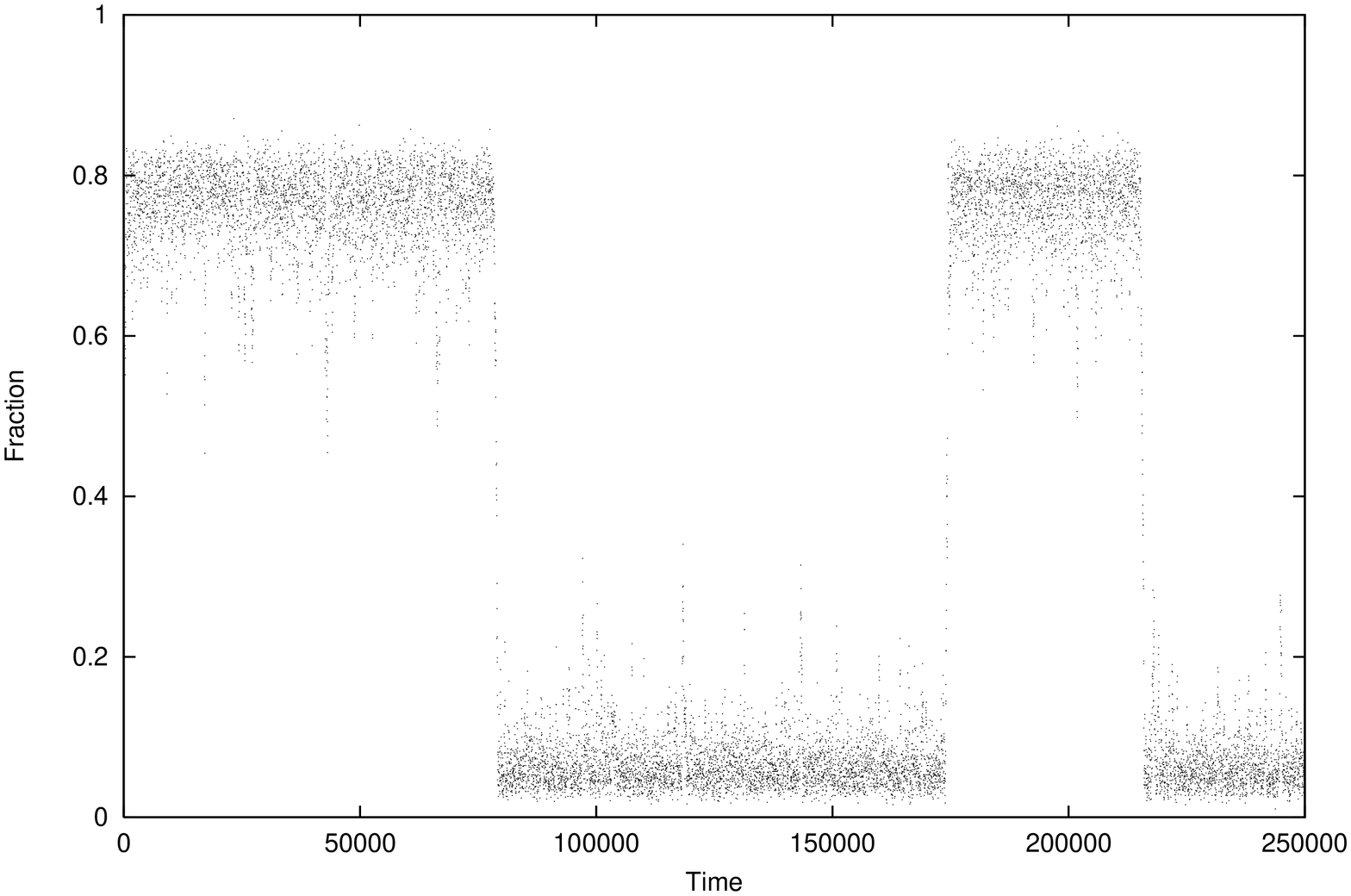}\\
\caption{\label{frequency3} Representative intervals of MC-generated time series of
the fractions of expectation types associated with the state space elements $-1.014$ (top),
$0.05$ (middle) and $1$ (bottom); the size of the population is 961 (31 $\times$ 31 square lattice), $\frac{1}{2J}=1.68$, $\delta=0$ 
}
\end{center}
\end{figure}

\end{document}